\documentclass[prl,a4paper,twocolumn,showpacs,groupedaddress,superscriptaddress]{revtex4}
\usepackage{graphicx}

\begin{document}

\title{Radiation Generated by Charge Migration Following Ionization}

\author{Alexander I. \surname{Kuleff}}
\email[e-mail: ]{alexander.kuleff@pci.uni-heidelberg.de}
\affiliation{Theoretische Chemie, Universit\"at Heidelberg, Im Neuenheimer Feld 229, 69120 Heidelberg, Germany}
\affiliation{Kavli Institute for Theoretical Physics, University of California, Santa Barbara, CA 93106-4030, USA}
%\altaffiliation[On leave from: ]{INRNE, BAS,72, Tzarigradsko Chaussee Blvd., 1784 Sofia, Bulgaria}
\author{Lorenz S. \surname{Cederbaum}}
%\email[Electronic mail: ]{lorenz.cederbaum@pci.uni-heidelberg.de}
\affiliation{Theoretische Chemie, Universit\"at Heidelberg, Im Neuenheimer Feld 229, 69120 Heidelberg, Germany}
\affiliation{Kavli Institute for Theoretical Physics, University of California, Santa Barbara, CA 93106-4030, USA}

\date{\today}

\begin{abstract}
Electronic many-body effects alone can be the driving force for an ultrafast migration of a positive charge created upon ionization of molecular systems. Here we show that this purely electronic phenomenon generates a characteristic IR radiation. The situation when the initial ionic wave packet is produced by a sudden removal of an electron is also studied. It is shown that in this case a much stronger UV emission is generated. This emission appears as an ultrafast response of the remaining electrons to the perturbation caused by the sudden ionization and as such is a universal phenomenon to be expected in every multielectron system.
\end{abstract}

\pacs{82.39.Jn, 33.20.-t, 42.65.Re}

\maketitle

If a multielectron system is ionized, a question of fundamental importance is how do the remaining electrons react. Whether the created hole remains unchanged for a certain time, or it changes its form and position as time proceeds due to the interaction with the remaining electrons, and, of course, what is the time scale of these changes. More than ten years ago it was shown \cite{CM_first} that after ionizing a molecule the many-body effects alone can beget rich ultrafast electron dynamics such that the hole created upon ionization can migrate throughout the system on the time scale of only few femtoseconds. Since the mediator of the process is the electron correlation and the electron relaxation, in order to distinguish it from the nuclear-dynamics-driven charge transfer, this phenomenon was termed charge migration. The charge migration after ionization of the valence shell has been theoretically studied in different molecules \cite{Breidbach03,Hennig05,dprop,Luennemann_JCP08}. In some specific, but not rare, cases the initially created hole localized on one end of the molecule was found to oscillate to the other end of the molecule and back within just a few femtoseconds. Being usually much faster than the nuclear motion, charge migration following ionization can thus be computed neglecting the nuclear dynamics, as long as one is interested in what happens within the time interval during which this ultrafast process takes place. Clearly, at later times the coupling to the nuclear dynamics will start to play a role and it has to be considered.  However, before the nuclear dynamics start to perturb the picture, the charge migration represents a charge oscillating throughout the molecule and, thus, the molecule itself can be seen as an oscillating dipole. It is well known that an oscillating dipole emits radiation. The natural question is then what is the emitted radiation and how strong it is.

In this Letter we show that the charge migration phenomenon generates a characteristic infrared (IR) radiation and compute the spectrum of this emission. In addition, we study the fundamental question what will happen if the ionization is performed extremely fast and obtain surprising results. We would like to note here that although it might appear academic, approaching the limit of sudden ionization is conceivable in view of the rapid development of the attosecond laser pulse techniques (see, e.g., Refs. \cite{Misha_atto,Giuseppe_atto}). 

The total power of the radiation emitted by a moving charge as a function of time can be calculated via the well-known Larmor formula (see, e.g., Ref. \cite{Jackson}), which in atomic units reads
\begin{equation}\label{power}
P(t) = \frac{2}{3c^3}|\ddot{\vec{D}}(t)|^2,
\end{equation}
where $\ddot{\vec{D}}(t)$ is the second time-derivative of the dipole moment
\begin{equation}\label{dipole}
\vec{D}(t) = \langle\Psi(t)|\hat{\vec{D}}|\Psi(t)\rangle,
\end{equation}
and $c$ is the speed of light. The spectrum of the emitted radiation can be obtained by substituting in Eq.~(\ref{power}) $\ddot{\vec{D}}(t)$ by its Fourier transform. Thus, in order to obtain the spectrum of the emission, one has to compute the time-dependent dipole moment of the cationic system. 

\begin{figure}[ht] 
\begin{center}
\includegraphics[width=7.5cm]{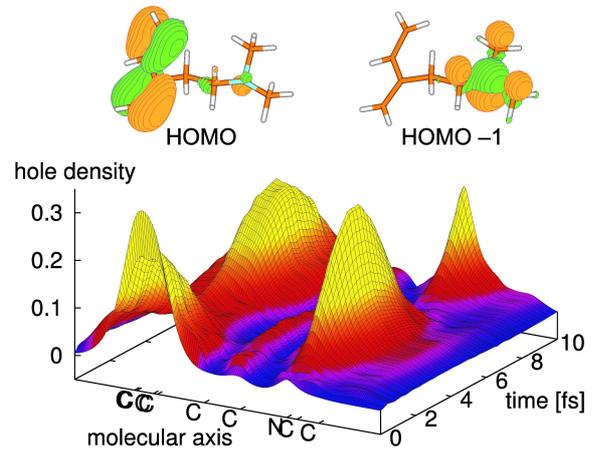}
\end{center}
\caption{\label{fig1}(color online) Hole density along the molecular axis of the molecule MePeNNA as a function of time after a localized ionization of the chromophore. The molecular axis is chosen to pass through the longest spatial extention of the molecule. The Hertree-Fock HOMO and HOMO$-1$ of the molecule are also shown.}
\end{figure}

Let us take a realistic show-case example, the molecule 3-methylen-4-penten-N,N-dimethylamine (MePeNNA), for which we showed in a recent work that after an outer-valence ionization a strong charge migration takes place \cite{Luennemann_JCP08}. Our many-body calculations showed that the outer-valence part of the ionization spectrum of the molecule consists of two ionic states, one at about 7.9~eV and one at about 8.4~eV (see Ref.~\cite{Luennemann_JCP08} and Fig.~\ref{fig3} below). These two states can be simultaneously populated via a laser pulse with photon energy centered between the two states and a band width sufficient to embrace both of them. Since the states are about 0.5~eV apart, one needs a pulse with duration of about 1~fs, ideally even shorter. Sub-femtosecond pulses are already available (see, e.g. \cite{Giuseppe2006}) and, thus, a simultaneous population of the two states is experimentally achievable. Of course, if the two ionization channels are open, they will interfere. Similar situation was realized experimentally and discussed in Refs.~\cite{Olga_Nature09,Olga_PNAS09}. Our calculations showed \cite{Luennemann_JCP08} that due to the many-body effects, each of the two ionic states is a strong mixture of two one-hole (1h) configurations, (HOMO)$^{-1}$ and (HOMO$-1$)$^{-1}$ (see also Fig.~\ref{fig3} below). The highest occupied molecular orbital (HOMO) of MePeNNA is localized on the chromophore moiety of the molecule, while the HOMO$-1$ on the N-terminal (see Fig.~\ref{fig1}). A constructive interference between the two ionization channels will lead to a localization of the initial hole on the chromophore, while a destructive interference will lead to a localization of the charge on the N-terminal. Let us take the ionic state prepared by the constructive interference, i.e. an initial hole localized on the chromophore, and see what will be the succeeding electron dynamics. Such an initial state can be prepared, e.g., by an intense positively chirped broadband laser pulse \cite{Cao_PRL98}, or by a $\pi$ pulse. For computing the charge migration we use elaborated many-body \textit{ab initio} methods only. A detailed description of the methodology can be found elsewhere \cite{Breidbach03,dprop}. 

The hole density of the molecule MePeNNA after creating the initial hole on the chromophore is shown in Fig.~\ref{fig1}. The hole density is given by the difference between the electronic densities of the system before and after the ionization \cite{CM_first,Breidbach03}.  It is clearly seen that immediately after the ionization, the hole starts to oscillate between the ``left'' and the ``right'' moiety of the system with a period of about 7.5~fs. These oscillations will be gradually distorted by the coupling to the nuclear degrees of freedom. The slower nuclear dynamics will eventually trap the charge on one of the two sites. However, the nuclear dynamics time scale is such that the hole will have time to perform a few, nearly perfect oscillations before the nuclear motion starts to perturb the picture. Thus, within the first few periods this charge migration process will represent an oscillating dipole.

\begin{figure}[ht] 
\begin{center}
\includegraphics[width=7.5cm]{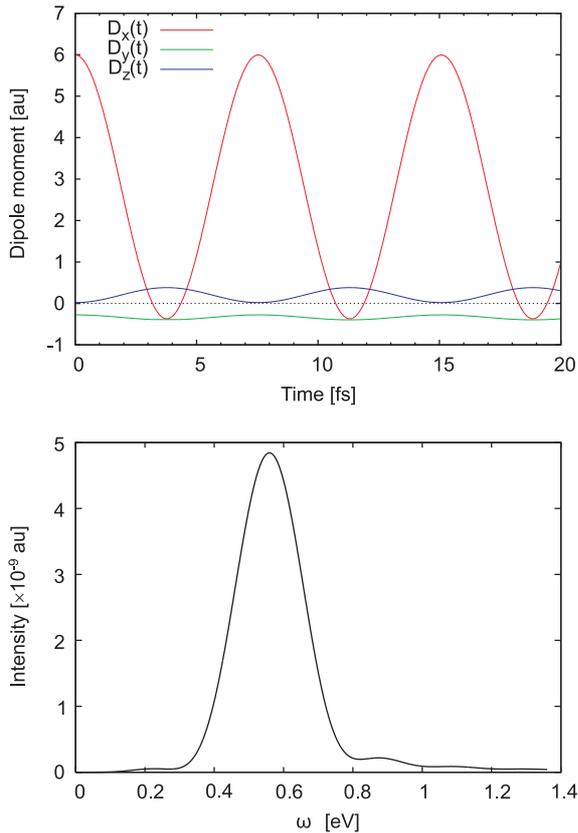}
%\vspace{10 cm}
\end{center}
\caption{\label{fig2}(color online) Components of the time-dependent dipole moment (upper panel) and the emission spectrum (lower panel) generated by the charge migration process initiated by a localized ionization of the chromophore site of the molecule MePeNNA.}
\end{figure}

In order to compute the time-dependent dipole moment of the cationic MePeNNA after the ionizing pulse is over, one can use the following, in principle exact, expansion of Eq.~(\ref{dipole}) 
\begin{equation}\label{dipole_exp}
\vec{D}(t) = \sum_{I,J} x_I^* \langle I |\hat{\vec{D}}|J\rangle x_J e^{-i\omega_{IJ}t},
\end{equation}
where $|I\rangle$ is a complete set of cationic eigenstates, $x_I=\langle I|\Phi_i\rangle$ is the transition amplitude with respect to the initially prepared non-stationary cationic state $|\Phi_i\rangle$, and $\omega_{IJ}=E_J-E_I$ is the difference between the corresponding cationic eigenenergies, or ionization potentials (IP). If the parameters of the ionizing pulse are chosen as discussed above, only the first two states will be populated and will give a non-zero contribution to the expansion (\ref{dipole_exp}). The components of the time-dependent dipole moment computed via Eq.~(\ref{dipole_exp}) are shown in the upper panel of Fig.~\ref{fig2}. By taking the second time-derivative of the dipole moment and Fourier transforming it, one obtains the spectrum of the emission generated by the oscillating charge shown in the lower panel of Fig.~\ref{fig2}. 

We see that the spectrum consists of a single peak at about 0.55 eV corresponding to the oscillation period of 7.5~fs. This emission is probably measurable. The total energy emitted during 10~fs radiation from a single molecule is $\sim 2.3\times10^{-9}$~eV. For density of $\sim 10^{18}$~molecules/cm$^3$, a typical density used in high-harmonic generation gas chambers \cite{Shan_APB02}, and interaction volume of $\sim 10^{-7}$~cm$^3$ \cite{Shan_APB02}, the charge migration will generate $\sim 115$ IR photons. Not long ago, the IR emission caused by the charge separation step of the phototransduction process of bacteriorhodopsin has been successfully measured \cite{JLMartin_PNAS04}. We would like to note that since, as we saw, the charge migration generates a characteristic radiation, the experimental observation of such an emission will be a direct proof for existence of the phenomenon of charge migration following ionization.

Let us now discuss the situation when the ionization is performed extremely fast, ideally infinitely fast. Of course, one cannot in practice remove an electron suddenly in the literal sense, but we have to note that attosecond laser pulses are already available and maybe in the not-so-distant future one will be able to approach this limit also experimentally in the following sense. It will suffice that the electron is removed such that during the process of ionization the other electrons do not have time to react. The time needed for the other electrons to respond to a sudden creation of a hole is about 50~asec \cite{Joerg_PRL05} (see also Ref.~\cite{tracing}). It was shown \cite{Joerg_PRL05} that this time is universal, i.e. it does not depend on the particular system, and as such appears as the time scale of the electron correlation. Thus, in practice, sudden ionization is equivalent to ionization performed faster than the electron correlation. If the ionization is performed faster than the electron correlation, one may assume that the electron is removed from a single molecular orbital, being a result of an independent particle model. The Hartree-Fock (HF) approximation provides the best independent particle theory to describe the ionization since the many-body corrections to the ionization energies begin to contribute at a higher order of perturbation theory than with other choices of approximations \cite{Szabo}. That is why, we may assume that the initial hole created upon sudden ionization is described favorably by a HF-orbital. On the other hand, if the sudden ionization has to be performed with a laser pulse it has to be extremely short, ideally shorter than 50~asec. We would like to note that isolated pulses with duration of about 80~asec are already available \cite{Goulielmakis_Sci2008}. Such pulse will have a very broad band allowing the population of a large number of the ionic states of the system. Thus, many ionic states $|I\rangle$ will contribute to the expansion (\ref{dipole_exp}). In addition, with such pulses one will not be able probably to address a particular HF orbital. In principle, all valence HF orbitals will participate and one will have to know how to separate the signal coming from the ionization out of the desired orbital. 

\begin{figure}[ht] 
\begin{center}
\includegraphics[width=7.5cm]{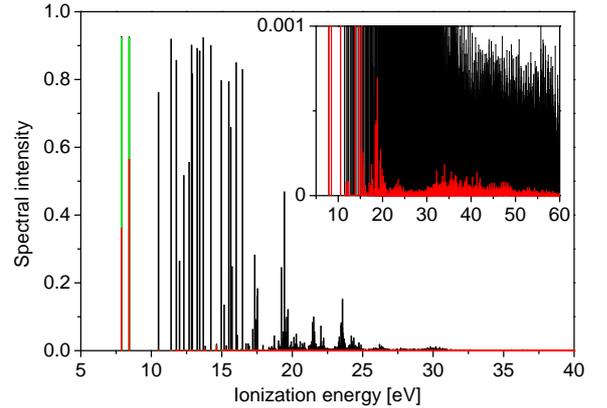}
%\vspace{5 cm}
\end{center}
\caption{\label{fig3}(color online) Ionization spectrum of the molecule MePeNNA computed using the \textit{ab initio} many-body Green's function method. The contributions of the 1h configuration (HOMO)$^{-1}$ to the cationic states is given in red. The contribution of the (HOMO$-1$)$^{-1}$ configuration in the first two ionic states is shown in green.}
\end{figure}

\begin{figure}[ht] 
\begin{center}
\includegraphics[width=7.5cm]{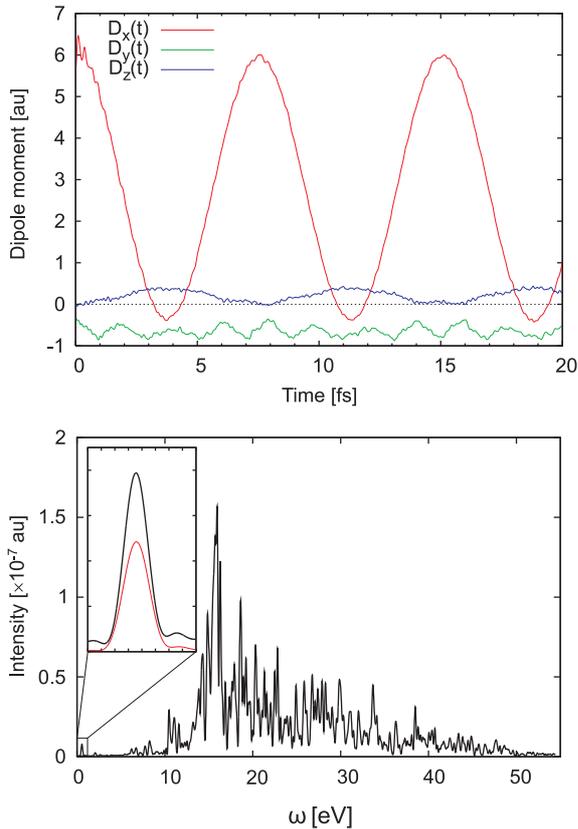}
%\vspace{10 cm}
\end{center}
\caption{\label{fig4}(color online) Components of the time-dependent dipole moment (upper panel) and the emission spectrum (lower panel) after sudden removal of an electron from the HF HOMO of the molecule MePeNNA. Comparison between the emission generated by suddenly removing an electron from the HOMO and the emission spectrum reported in Fig.~\ref{fig2} is shown in the inset.}
\end{figure}

Let us assume, for the moment, that one can remove an electron from a particular HF-orbital and see what will happen. As an example we will take again the molecule MePeNNA and will remove an electron from its HOMO, which, as we discussed above, is localized on the chromophore (see Fig.~\ref{fig1}). The ionization spectrum of the molecule, calculated via Green's function methodology \cite{non-Dyson}, is shown in Fig.~\ref{fig3}. The (HOMO)$^{-1}$ contributions to the ionic states are shown in red in the figure. We see that apart from the two states already discussed above, the configuration (HOMO)$^{-1}$ contributes also to a large number of ionic states spread over large energy range (see the inset of Fig.~\ref{fig3}). Thus, a sudden removal of an electron from the HF HOMO will create a superposition of a large number of ionic states weighted by the corresponding transition matrix elements. The time-dependent dipole moment created by such initial wave packet is shown in the upper panel of Fig.~\ref{fig4}.

We see that apart from the strong oscillation with a period of about 7.5~fs, the components of the dipole moment show small wiggles on a time scale of only few tens of attoseconds. These, of course, come from the coupling between states separated by few tens of eV. Although small, these variations of the dipole moment generate a strong emission in the ultraviolate (UV) range. This can be clearly seen in the lower panel of Fig.~\ref{fig4}, where the emission generated by the charge migration following sudden ionization of the HF HOMO of the molecule is shown. We see that the spectrum spreads up to about 50~eV with the most intense emission in the range 10--25~eV. The intensity of this UV emission is two orders of magnitude stronger than the IR radiation produced by the 7.5~fs oscillations of the charge between the two ends of the molecule. The latter is shown in the inset of the lower panel of Fig.~\ref{fig4} together with the emission produced by an initial state constructed as a superposition of the two states bearing the largest contribution of the configuration (HOMO)$^{-1}$, discussed above and shown in Fig.~\ref{fig2}. Interestingly, the white noise generated by the contribution of a large number of ionic states introducing a superposition of a large number of different frequencies, gives rise to a substantial increase of the IR emission produced by the 7.5~fs charge oscillation.

We have to emphasize that the UV emission generated by the attosecond variations of the hole charge is not inherent only to the cases when a strong charge migration within the molecule is observed. These variations appear as a result of the electronic correlation present in the cation and as such will accompany any sudden ionization of a manyelectron system, irrespective of the succeeding dynamics. Removing an electron from each valence HF orbital will lead to an emission in the UV range similar to the one shown in Fig.~\ref{fig4}. Thus, exposing the system to an ultrashort pulse will generate an UV emission which will be a superposition of the emission spectra coming from the ionization of each individual HF orbital.

Let us conclude. We demonstrated that the purely electronic phenomenon of ultrafast charge migration following ionization of a molecular system generates a characteristic radiation. Since the charge migration typically represents charge oscillations with a few femtoseconds period, the generated emission is in the IR range. Encouraged by the successfull measurement of the emission produced by the nuclear-dynamics-driven charge transfer in bacteriorhodopsin \cite{JLMartin_PNAS04}, we believe that the emission generated by a purely electron-correlation-driven charge migration could be measurable. That is why we advocate here experimental efforts in this direction. Due to the fact that the charge migration produces a characteristic radiation, the observation of such IR emission would be a direct experimental proof for the charge migration phenomenon. We also studied the situation when the initial ionic wave packet is produced by a sudden removal of an electron and found out that in this case a much stronger UV emission is generated. Moreover, this emission appears as an ultrafast response of the remaining electrons to the perturbation caused by the sudden ionization and as such is a universal phenomenon to be expected in every multielectron system irrespective of the succeeding electron dynamics. 

We hope that our results will trigger future experimental studies on the fascinating subject of ultrafast charge migration and ultrafast ionization.

The authors thank Andr\'e Bandrauk for enlightening discussions. This research was supported in part by the National Science Foundation under Grant No. NSF PHY05-51164. Financial support by the DFG is gratefully acknowledged.

\end{document}